\documentclass[twocolumn,preprintnumbers,amssymb,amsmath,superscriptaddress,letterpaper,nofootinbib]{revtex4}[12pt]

\usepackage{graphicx}
\usepackage{dcolumn}
\usepackage{bm}
\usepackage{natbib}
\usepackage{pstricks}
\usepackage{epsfig}
\usepackage{epstopdf}
\usepackage{mathtools}
\usepackage{slashed}
\usepackage{physics}
\usepackage{pgf} 

\newcommand{\be}{\begin{equation}}
	\newcommand{\ee}{\end{equation}}
\newcommand{\bea}{\begin{eqnarray}}
	\newcommand{\eea}{\end{eqnarray}}

\begin{document}

	\title{Spin-Graviton-Spin: a unique probe of quantum gravity}

	\author{Radmehr Fathi}
	\email{radmehr.fathi82@sharif.edu}
	\affiliation{Department of Physics, Sharif University of Technology, Tehran 11155-9161, Iran}
	
	\author{Nima Khosravi}
	\email{nima@sharif.edu}
	\affiliation{Department of Physics, Sharif University of Technology, Tehran 11155-9161, Iran}
	

	\begin{abstract}
		The existence of quantum gravity is the most important question in theoretical physics. Besides the theoretical development in this direction, looking for its low-energy consequences can provide clues. However, besides availability, it is crucial to ensure that what is observed uniquely comes from quantum gravity and not from other forces in the environment. What is the unique feature of gravity? The spin-$2$ nature of its propagator. In this Letter, we calculate what the unique feature of the spin-$2$ graviton is in the interaction of two spin-$\frac{1}{2}$ particles. This unique feature shows itself in intertwining the spin and momentum of two particles simultaneously as a dipole-dipole interaction. This cannot happen for any spin-$0$ or spin-$1$ propagator, at least at leading order, which scales as $\frac{1}{r^3}$ in real space. We re-calculate the results using an independent approach, i.e., the Foldy-Wouthuysen transformation, and reach the same results. Due to simultaneous coupling of both particles' spins and momenta, a four-partite entanglement framework is suggested to observe this effect.

	\end{abstract}
	
	\maketitle

	\section{Introduction}
	It is not exaggerated if one says the quest for the intertwining of quantum and gravity is the most profound question in theoretical physics. This quest has been around for over a century, with much effort from giants. The main issue is that the (full non-linear) Einstein gravity is not renormalizable\footnote{There is a higher-order gravity model which is renormalizable, though it suffers from the ghosts \cite{PhysRevD.16.953}.}. Though there are some theoretical suggestions, such as string theory \cite{Polchinski:1998rq}, loop quantum gravity \cite{Ashtekar:2004eh}, etc. They suffer from either theoretical inconsistency or a lack of observational support. The latter is difficult because, in principle, detecting any quantumness of gravity requires very high-energy probes. These unsuccessful attempts have even led to a paradigm shift suggesting that gravity need not be quantum, though no reliable model supports this claim. Recently, there has been an interesting idea that one can observe the effects of quantum gravity at low energy scales by looking for gravitationally induced entanglement. This approach is based on the LOCC principle, which states that local operations and classical communications cannot create entanglement. This approach makes the experimental setups more accessible.
	In this direction, there is a proposal on entanglement of two masses \cite{PhysRevLett.119.240402,PhysRevLett.119.240401}, known as the quantum gravity induced entanglement of matter experiment (QGEM). The QGEM suggests looking for entanglement between two small masses while Newtonian gravity is in action. If there is any entanglement, then gravity must be quantum since LOCC does not allow a classical interaction to create entanglement. There have been some discussions regarding what can definitively be concluded from observing gravitationally induced entanglement; see \cite{PhysRevD.108.L101702}. Since gravity is extremely weak at these scales, observing its outcome requires making the setup very clean from any other (external) fields. This is important because the proposed measure is based on Newton's inverse-square law. This part of the gravitational force is due to the scalar part of the metric which appears in the gravitational propagator. Though the scalar part is not a propagating degree of freedom of Einstein gravity, it shows itself off-shell. This scalar part can be mimicked by other fields\footnote{For example, Coulomb's law also has the same form, and if there is a small electrical charge, then it may produce the same feature, and it might not be straightforward to distinguish its effect from gravity’s effect.}.  This can be a motivation to look for a unique feature in gravitational quantum effects. Some of the many works on quantum information effects in quantum gravity can be found in \cite{PhysRevLett.119.240402,PhysRevLett.119.240401,marletto2024quantuminformationmethodsquantumgravity,PhysRevA.101.052110,PhysRevD.108.L101702,Krisnanda_2020,Marletto_2018,Bose_2022,Christodoulou_2023,Chakraborty_2023,Beckering_Vinckers_2023} while they are all focusing on the gravitational interactions that can be mimicked by other forces.

	The goal of the present work is to find a unique effect of the Einstein gravity (and perhaps all the valid gravitational models). The unique property of any gravitational field theory is that its propagator should be spin-$2$ or at least have it in its degrees of freedom, see also \cite{vedral2022realproblemsquantumgravity}. It is well-known that the spin-$0$ cannot be coupled to radiation, which breaks the universality of gravity, and the spin-1 makes (effectively) positive and negative masses, which is against observations (and always attractiveness of gravity). Mathematically, a spin-$2$ particle (the graviton) is the traceless-transverse part of a field (the metric). In order to derive this unique interaction, we will be examining the gravitational interaction between two spin-$\frac{1}{2}$ particles. Some aspects of the gravitational interaction between two spin-$\frac{1}{2}$ particles can be found in \cite{PhysRevD.102.084015,campos2022investigationseffectiveelectromagneticgravitational,majumder2026xirphi2nonminimalcoupling,holstein2008spineffectslongrange,Obukhov_2002,PhysRevD.67.084033,PhysRevD.2.1428,Barker:1966zz,PhysRevD.12.329,Beckering_Vinckers_2025}. If one could find such interactions, then they could be used as a unique footprint of a quantum theory of gravity e.g. in tabletop experiments. As expected, the amplitude of the unique feature will be tiny even compared to other gravitational interactions. So though it is theoretically important but it may not be achievable in tabletop experiments with simple setups. This may hint at more complicated tabletop experiments, e.g., containing a huge number of particles or focusing on the signals from the skies where strong gravitational fields are available, e.g., near black holes. This is the price that we have to pay to look for the unique footprint of a quantum gravity model. 
	
	\section{Graviton}
	The linear GR in the Minkowski background, $g_{\mu\nu}=\eta_{\mu\nu}+h_{\mu\nu}$, can be formulated as 
	\begin{eqnarray}
		L= h_{\mu\nu}{\cal{E}}^{\mu\nu\alpha\beta}h_{\alpha\beta}+\kappa h_{\mu\nu}T^{\mu\nu}
	\end{eqnarray}
	where $\kappa$ the coupling constant is related to the Newton constant $\kappa = 8 \pi G$ and $T^{\mu\nu}$ is the energy-momentum tensor.
	This can be written as
	$
	\Box{\overline{h}_{\mu\nu}}=-2\kappa T_{\mu\nu}
	$ 
	where we have implemented the Lorenz gauge, $\partial^\mu h_{\mu\nu}=1/2 \partial_\nu h $, and defined $\overline{h}_{\mu\nu}=h_{\mu\nu}-1/2\, h\,\eta_{\mu\nu}$ to make the calculations easier without loss of any generality. In the following, we will use the above equation in the Fourier space\footnote{Note that we do not use a new symbol for the fields in the Fourier space.} $k^2 \overline{h}_{\mu\nu}=-2\kappa T_{\mu\nu}$. Another result from the above Lagrangian
	is the form of the graviton propagator 
	\begin{eqnarray}\label{propagator}
		D_{\mu \nu \rho \sigma}=\frac{i}{2 k^2}(\eta_{\mu \rho} \eta_{\nu \sigma}+\eta_{\rho \nu} \eta_{\mu \sigma}-\eta_{\mu \nu}\eta_{\rho \sigma}).
	\end{eqnarray}
	So the scattering of the first particle from the second one, is given by:
	\begin{eqnarray}\label{eq:scat}
		i\mathcal{M}=\frac{- \kappa}{4}(T_1^{\mu \nu} D_{\mu \nu \rho \sigma} T_2^{\rho \sigma})=	\frac{-i \kappa}{4 k^2}(T_{1 \mu \nu} T_2^{\mu \nu}-\frac{1}{2}T_1 T_2)
	\end{eqnarray}
	where sub-indices $1$ and $2$ represent the first and the second particles, respectively. The equality is given by applying (\ref{propagator}) and $T=\eta^{\mu\nu}T_{\mu\nu}$.
	
	Since our focus is on the traceless-transverse (TT) part of $h_{\mu\nu}$, which represents the two degrees of freedom of a massless spin-$2$ graviton, we need to take this TT part out of the equations. First, we introduce a projection operator which can do it for any general tensor $\Lambda_{i j k l} A^{ij}=A^{TT}_{kl}$ in which
	\begin{eqnarray}
		\Lambda_{i j k l}= P_{i k} P_{j l}-\frac{1}{2} P_{i j} P_{k l},
	\end{eqnarray}
	where $P_{i j}= \delta_{i j}-\frac{k_i k_j}{k^2}$ and $\Lambda_{i j k l} \Lambda^{i j m n} =\Lambda_{k l}^{\vspace{0 cm}m n}$.
	We replaced the spacetime indices, $\mu$'s, by the spatial indices, $i$'s, since the $h^{TT}_{ij}$ part is the graviton. The scattering (\ref{eq:scat}) (without the coefficient) for the TT part can be written as 
	\begin{eqnarray}\label{scat2}
		T^{ij}_1 D^{TT}_{ijkl} T^{kl}_2&=&T_1^{i j}\Lambda_{ij}^{mn} D_{mnpq} \Lambda^{pq}_{kl} T_2^{kl}	 =\frac{i}{k^2}\big[T^{ij}_1\Lambda_{ijkl} T^{kl}_2 \big] \nonumber \\
		&& \,\
	\end{eqnarray}

	\subsection{Spin-$\frac{1}{2}$}
	Now it is time to turn on the matter field. For our purposes, we focus on spin-$\frac{1}{2}$ particles. The corresponding energy-momentum tensor\footnote{Note that it is the zeroth-order energy-momentum tensor in terms of metric (tetrads). Since we want to keep all the equations at first order of the metric, then the matter field should be at zeroth order. Otherwise, the higher-order terms in $T_{\mu\nu}$ pump second-order and higher-order terms via the Einstein equation.} is \cite{Parker:2009uva,Birrell:1982ix}:
	\begin{eqnarray}
		T_{\mu \nu}&=&\frac{i}{2} \biggl[ \bar{\psi} \gamma_\mu \partial_\nu \psi + \bar{\psi} \gamma_\nu \partial_\mu \psi  - \partial_\mu \bar{\psi} \gamma_\nu \psi - \partial_\nu \bar{\psi} \gamma_\mu \psi \biggr] \nonumber \\
		&& \,\
	\end{eqnarray}
	where $\psi$ is the spinor fields and $\gamma$'s are Dirac matrices. The $ij$'s part of the energy-momentum tensor in the Fourier space will be
	\begin{eqnarray}\label{spin-emt}
		T_{ij}(k)&=& \frac{1}{2} \int \frac{d^3 p}{(2\pi)^3}\bigg[ (2 p_j-k_j) \bar{\psi}_{(p-k)} \gamma_i \psi_{(p)} \nonumber \\
		&+& (2 p_i-k_i) \bar{\psi}_{(p-k)} \gamma_j \psi_{(p)} \bigg]
	\end{eqnarray}

	\section{Spin-Graviton-Spin}
	Now, by substituting the above relation (\ref{spin-emt}) into (\ref{scat2}), we can calculate the scattering of two spin-$\frac{1}{2}$ particles due to a spin-$2$ graviton.
	But the interesting part for our purposes is the terms where both spins and momenta appear simultaneously. These terms are
	\begin{eqnarray}\label{eq:TT}
		&& \frac{i}{2 k^2} \bigg[ \Big( (2\vec{p}_1-\vec{k}) \cdot \vec{\alpha}_2 \Big)
		\Big( (2\vec{p}_2-\vec{k}) \cdot \vec{\alpha}_1 \Big) 
		\\ \nonumber
		&& + (\vec{\alpha}_1 \cdot \vec{\alpha}_2) ((2\vec{p}_1-\vec{k}) \cdot (2\vec{p}_2-\vec{k})) \bigg] \subset T^{ij}_1 D^{TT}_{ijkl} T^{kl}_2
	\end{eqnarray}
	where $\alpha_i=\bar{\psi}_{(p-k)} \gamma^i \psi_{(p)}$. We note that we are only keeping the terms that contain a direct coupling of the indices of the first and second particle in both the spin and momentum. The reason for this will be clear in the next section. If we take the non-relativistic limit, we would get $\alpha_i=\frac{\chi_{(p-k)}}{2 m}[2 p_i -k_i+i (\vec{k} \times \vec{\sigma})_i] \chi_{(p)}$ where $\chi$ is the matter part of the wave function in $\psi=\binom{\chi}{\xi}$. Now, if we take the Fourier transformation and present the above result in coordinate space, we will get the effective Hamiltonian between our particles:
	\begin{eqnarray}\label{eq:HTT}
		H^{\text{spin}-2}=H^{\text{spin}-2}_{\text{mag.}}+H^{\text{spin}-2}_{\text{elec.}}
	\end{eqnarray}
	where 
		\begin{eqnarray}\label{eq:dipole1}
	H^{\text{spin}-2}_{\text{mag.}}&=&	\frac{1}{r^3} \Big( 3(\hat{r} \cdot \vec{\mu}_1) (\hat{r} \cdot \vec{\mu}_2) - (\vec{\mu}_1 \cdot \vec{\mu}_2)\Big) \nonumber \\
	&+&\frac{8 \pi}{3} (\vec{\mu}_1 \cdot \vec{\mu}_2) \delta^3 (r), 
	\end{eqnarray}
is the dipole-dipole interaction for the magnetic-like part if
	\begin{eqnarray}\label{dipole-2}
		\vec{\mu}_1=\vec{\sigma}_1  (\vec{\pi}_1 \cdot \vec{\pi}_2)^{1/2} \hspace{1cm} \text{and}\hspace{1cm}
		\vec{\mu}_2=\vec{\sigma}_2 (\vec{\pi}_1 \cdot \vec{\pi}_2)^{1/2}. \nonumber
	\end{eqnarray}
	and the electric-like part	is given by
	\begin{eqnarray}\label{eq:dipole2}
	H^{\text{spin}-2}_{\text{elec.}}&=&	\frac{1}{r^3} \Big( 3(\hat{r} \cdot \vec{\mu}_1) (\hat{r} \cdot \vec{\mu}_2) - (\vec{\mu}_1 \cdot \vec{\mu}_2)\Big) \nonumber \\
	&-&\frac{4 \pi}{3} (\vec{\mu}_1 \cdot \vec{\mu}_2) \delta^3 (r),
	\end{eqnarray}
	where
	\begin{eqnarray}\label{dipole-1}
		\vec{\mu}_1=\vec{\sigma}_1 \times \vec{\pi}_2 \hspace{1cm} \text{and}\hspace{1cm}
		\vec{\mu}_2=\vec{\sigma}_2 \times \vec{\pi}_1 \nonumber.
	\end{eqnarray}
	
	Equation (\ref{eq:HTT}) comes from the $T_1^{\mu \nu} T_{2\mu \nu}$ term in the scattering amplitude. It is also part of the TT part of the scattering amplitude. This states that the graviton itself is responsible for this interaction. The above result could be written, very interestingly, as a sum of magnetic-like and electric-like dipole interactions. 

	This is the main result of this work: exchanging a spin-$2$ particle causes the coupling between the spin of one particle and the momentum of the other one(and vice versa) in the form of dipole-dipole interactions. 
	\section{Uniqueness}
	The claim is that the above result is a unique feature of a spin-$2$ propagator. The reason is that in spin-$1$ theories such as QED, the scattering amplitude has the form $j_1^{\mu} D_{\mu \nu} j_2^{\nu}$, where in the Feynman gauge, $D_{\mu \nu} (k)=\frac{- g_{\mu \nu}}{k^2}$. This results in a scattering amplitude of the form $i\mathcal{M}=\frac{- j_1^\mu j_{2 \mu}}{k^2}$. In this term, the density current has only one index, and both particles' density currents have the same indices. At the tree-level diagrams, couplings with the form of (\ref{eq:HTT}) are not possible with a spin-$1$ propagator since it requires the coupling of two different indices of the density currents in the scattering amplitude. For this reason, the type of couplings that appear in equation (\ref{eq:HTT}) is a unique signature of a spin-$2$ theory. It is worth noting that if we add higher-order terms like $A_\mu A_\nu J^\mu J^\nu$ in the Lagrangian, we would get couplings like (\ref{eq:HTT}), though such a term is not gauge invariant.  At higher orders, i.e., loop diagrams in QED such as box or cross diagrams, couplings in the same format as (\ref{eq:HTT}) are possible. This is because loop diagrams contain two photons (i.e., terms like $A_\mu A_\nu $ appear), which can make up for the missing index and induce those types of couplings. However, because they are higher-order correction terms, they cannot produce the same type of interactions at scale $\frac{1}{r^3}$, especially dipole-dipole terms of that form.  So in conclusion, this Hamiltonian is a unique signature of spin-2 theories such as gravity. 
	This result is crucial and useful when designing an experimental setup to observe the effects of quantum gravity. This helps ensure there is no leakage from other particles, and that the observed signal is due only to a graviton.

	\section{An alternative approach: The Foldy-Wouthuysen  transformation}
	The Foldy-Wouthuysen (FW) transformation is a procedure to take the non-relativistic limit of the Dirac equation. This transformation in the case of the electromagnetism \cite{PhysRev.78.29,PhysRev.111.1011,Silenko_2003,Murgu_a_2010,Wienczek_2022} and also for the gravity \cite{PhysRevD.88.084014,PhysRevD.84.024025,PhysRevD.80.064044,PhysRevD.76.061101,PhysRevD.82.104056,alcubierre2025diracequationgeneralrelativity} is fully studied in the literature. For our purposes, the FW transformation gives the same results as what we have already discussed, but it is useful to report it here: first, to confirm our previous results with an independent alternative procedure, and second, because the FW procedure makes it easier to handle additional terms.
	
	The result of the FW transformation for a Dirac particle in a curved spacetime will be:
	\begin{eqnarray}\label{eq:FW-curved}
		E \chi &=& 
		\bigg( m (1-\frac{h_{00}}{2}) + \Phi_g + \frac{1}{2}(\vec{\Gamma} \cdot \vec{\sigma}) \nonumber \\
		&+&\frac{(\vec{\Omega} \cdot \vec{\sigma})^2}{2m}+\frac{1}{2 m} [\beta \vec{\pi} \cdot \vec{\alpha}, \mathcal{E}]_{\text{matter}}\bigg) \chi
	\end{eqnarray}
	where $\chi$ is the matter part of the wave function in $\psi=\binom{\chi}{\xi}$. In the linearized regime we have:
	\begin{eqnarray}
		\Gamma^l &=&\frac{1}{2} (\partial_k h_{j0}) \epsilon^{kjl} \\
		\Phi_g &=&  h^i_{\ \ 0} \pi_i - \frac{i}{2} ( \partial_j h^{0j} - \partial_0 h^j_{\ \ j}) \\
		\Omega_l&=&\left( 1 - \frac{h_{00}}{2} \right) \pi_l+ \frac{1}{2} h^i_{\ \ l} \pi_i \nonumber \\
		&-& \frac{i}{2} \partial_l (h_{00}) -\frac{i}{2} (\partial^i h_{il} - \partial_l h^i_{i}).
	\end{eqnarray} 
	The details of the calculations for the FW transformation can be found in Appendix \ref{app-FW}. Now, if we use a source-probe approach as in \cite{Paper-QED-Spin-Spin} and find the $h_{0 i}$ induced by the second particle in the location of the first particle, and plug it into the $\vec{\Gamma}\cdot \vec{\sigma}$ term, we would get a Breit-like interaction \cite{PhysRevD.2.1428}:
	\begin{eqnarray}
		\frac{G}{4} \left( \frac{3(\sigma^{(1)} \cdot \hat{r})(\sigma^{(2)} \cdot \hat{r})}{r^3} - \frac{\sigma^{(1)} \cdot \sigma^{(2)}}{r^3}+ \frac{8 \pi}{3} (\sigma^{(1)} \cdot \sigma^{(2)}) \delta(r) \right). \nonumber
	\end{eqnarray}
	For the derivation of similar interactions, see \cite{PhysRevD.2.1428,Barker:1966zz,PhysRevD.12.329}. As expected, this interaction has the same form as the Breit interaction \cite{PhysRev.34.553,PhysRev.39.616}. This is because we are dealing with $h_{0 i}$, which is the analog of the magnetic vector potential in gravitomagnetism \cite{mashhoon2008gravitoelectromagnetismbriefreview}. Since we are looking for unique terms for gravity, we need to examine terms that contain $h^{TT}_{i j}$. We can find such terms in the $(\vec{\Omega} \cdot \vec{\sigma})^2$ part of the FW transformation. 
	We are only interested in terms that contain both the spins and momentum of our particle, so we keep only the relevant terms:
	\begin{eqnarray}\label{eq:sym}
		 \frac{1}{4m}i \epsilon_{ijk}  \sigma^k  \Bigl[\left( \pi^i h^{lj} \pi_l \right)
		+ (h^{lj} \pi^i \pi_l + h^{il} \pi_l \pi^j)\Bigr] \chi.
	\end{eqnarray}
	We note that if we assume that our particles do not have any electromagnetic interactions, the second parenthesis in (\ref{eq:sym}) vanishes. In order to derive (\ref{eq:HTT}), we will use the same procedure as in \cite{Paper-QED-Spin-Spin}. We first assume that there is no electromagnetic interaction between our particles. \\
	If we solve the Einstein field equations for $h_{i j}$ using the non-relativistic limit, we would get (the details of these calculations can be found in Appendix \ref{app-fieldeq}):
	\begin{eqnarray}\label{eq:hij}
		\overline{h}_{ij} = \frac{\kappa}{8\pi m} 
		\int \chi^{\dagger} \left( \frac{r^k}{r^3} \right)\sigma^m\biggl[ \pi_j \epsilon_{m k i} 
		+\pi_i \epsilon_{m k j} 
		\Biggr]\chi d^3x 
	\end{eqnarray}
	where $\vec{r}$ is the difference vector of our particles. Now, if we use (\ref{eq:hij}) in the first term of (\ref{eq:sym}), we can get the same Hamiltonian as in (\ref{eq:HTT}). This is our unique spin-2 Hamiltonian.
	
	 \subsection{Spin-Graviton-Magnetic field coupling}
	 We can also derive another interaction between spin, gravity, and the magnetic field using the second term in (\ref{eq:sym}). If we assume that our spin-$\frac{1}{2}$ particles have electromagnetic interactions. We can simplify the second term in (\ref{eq:sym}):
	\begin{eqnarray}\label{eq:ssgB}
		\frac{i \epsilon_{ijk}}{2} (h^{lj} \pi^i \pi_l + h^{il} \pi_l \pi^j)\sigma^k=
		\frac{q}{2} (h^{k l}B_l\sigma^k-h \vec{B} \cdot \vec{\sigma})
	\end{eqnarray}
	where $q$ is the electric charge of the particle and $\vec{B}$ is the magnetic field. In this equation, if we assume that we have an external magnetic field and also assume that the second particle induces the metric perturbations, we would have:
	\begin{eqnarray}
		&& \frac{\kappa q}{16 \pi m r^3} \big[ ((\vec{\sigma}_2 \times \vec{r})\cdot \vec{B} (r_1)) (\vec{\pi}_2 \cdot \vec{\sigma}_1) \nonumber \\
		&& + ((\vec{\sigma}_2 \times \vec{\sigma}_1)\cdot \vec{r}) (\vec{\pi}_2 \cdot \vec{B} (r_1))\big]
	\end{eqnarray}
	where $r_1$ is the location of the first particle. We keep only the first term on the RHS of equation (\ref{eq:ssgB}). This interaction is interesting for multiple reasons. First, it shows that the spin and momentum of the second particle affect the interaction between the magnetic field and the first particle's spin, and this is possible because of gravity. To observe this effect, in principle we could increase the external magnetic field until the interaction strength is strong enough to be observed. However, doing so would probably make the other electromagnetic interactions so strong that they would immediately destroy the system's quantumness. The magnetic field amplitude we would need is also not achievable with current technology. We note that if we assume that both of our particles have a quantum feature, then we should also add a similar term for the second particle as well. The FW transformation may be more complicated than the scattering amplitude approach. However, because it handles all terms at once, the FW approach makes it more straightforward to derive the spin-graviton-magnetic field coupling terms.
	\section{Gravitationally induced multi-partite entanglement}
\begin{figure}[t]
		\centering
		\resizebox{\columnwidth}{!}{\input{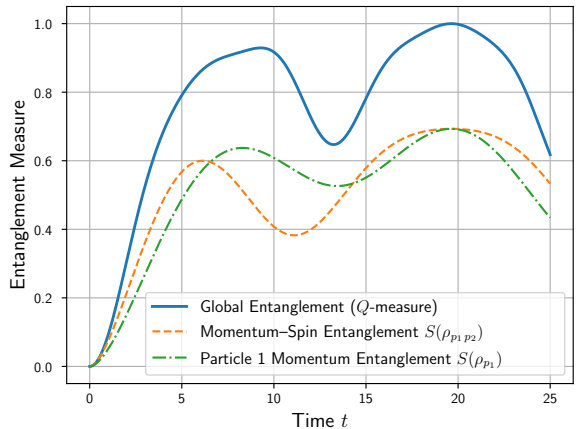}}
		\caption{The Meyer-Wallach Q-measure for global entanglement and the von Neumann entropy of the momentum-spin entanglement and the first particle momenta. For drawing this graph we assumed that $J= 1 (\text{eV})$ and $\frac{p_0}{m}=0.2,\frac{p_1}{m}=\frac{p_2}{m}=0.1$.}
		\label{fig:QSS}
	\end{figure}

	 In our case, since we have an interaction between the spins and the momenta of both of our particles, we can explore the possibility of gravitationally induced multi-partite entanglement. To do this, we will assume two spin-$\frac{1}{2}$ particles moving in the same direction with a constant distance of $r$. We assume that the only interaction present between the particles is (\ref{eq:HTT}). This assumption is only for a toy-model examination, since even in the best scenarios, Newtonian and Breit-like gravitational interactions will be present. We assume that the direction of the momentum of the particles is the $z$ direction and the direction of the distance is the $y$ direction. In this scenario, equation (\ref{eq:HTT}) will look like this:
	\begin{eqnarray}
		H=\frac{- J (\pi_1 \otimes \pi_2)}{m ^2} \biggl[\sigma_x \otimes \sigma_x-\sigma_y \otimes \sigma_y+\frac{1}{3}\sigma_z \otimes \sigma_z \biggr]
	\end{eqnarray}
	where $J=\frac{3 \kappa}{32 \pi r^3}$. For simplicity, we assume that the distance is a fixed quantity and the momenta are two-state systems. We also assume that the operator for the momenta has the form:
	\begin{eqnarray}
		\pi_1=
		\begin{pmatrix}
			p_0+p_1 & 0 \\
			0 & p_0-p_1
		\end{pmatrix}
		, 
		\pi_2=
		\begin{pmatrix}
			p_0+p_2 & 0 \\
			0 & p_0-p_2
		\end{pmatrix}
	\end{eqnarray}
	where $p_0,p_1,p_2$ are fixed parameters. Now, suppose that the system is in the unentangled state $\ket{++} \ket{00}$, (where the first two qubits are for the momentum of the first and second particles and the third and fourth qubits are for the spin of the first and second particles) we can compute the Meyer-Wallach Q-measure for global entanglement in the four-partite system, as well as the von Neumann entropy of chosen bipartitions. The Meyer-Wallach Q-measure for global entanglement of an N-qubit pure state is \cite{Meyer_2002}:
	\begin{eqnarray}
		Q=2(1-\frac{1}{N} \sum_k \Tr{\rho_k^2}).
	\end{eqnarray}
	The value of the Q-measure and the von Neumann entropy of this system can be seen in Figure \ref{fig:QSS}. Obviously, if we use feasible values for $J$, we would get extremely small values for the entanglement measures. This is something we would expect, since gravity itself is really weak and we are probing it with spin-$\frac{1}{2}$ particles whose mutual interaction is itself tiny. To extract a unique feature, we are using the next-order term in the non-relativistic limit. This makes the effect extremely small even compared to usual gravitational interactions. This is the price to pay for uniqueness. The toy model we study may be impractical in the laboratory, but it may still be useful for suggesting ways to design experiments that target this feature in the future. The amplitude of the unique feature might become more accessible in curved spacetime backgrounds. For example, if we examine this effect near a rotating black hole, we would expect the amplitude to become more accessible. This suggests that while laboratory detection is difficult, the effect may eventually become accessible in cosmic experiments.

	\section{Conclusions and Discussions}
	We have derived the gravitational interaction between the spin and momentum of two spin-$\frac{1}{2}$ particles. Two independent approaches are studied: the one-graviton exchange and the FW transformation. We showed that this interaction can be a \textit{unique} probe of quantum gravity. The uniqueness is based on the traceless-transverse property of a spin-$2$ particle, which makes the graviton distinguishable from other fields. 
	Theoretically, other spin-$2$ fields (e.g., spin-$2$ dark matter) can produce the same effects. 
	This uniqueness offers a distinct theoretical advantage for tabletop quantum gravity proposals, such as those based on the (QGEM) protocol. One issue with QGEM-like proposals is their focus on the Newtonian potential part of Einstein gravity. This part of Einstein gravity comes from the (off-shell) scalar degree of freedom of the model and can be mimicked by any other scalar field. So it is important to focus on the spin-$2$ part of the degrees of freedom to ensure the induced entanglement is purely gravitational. However, the strength of the unique spin-$2$ interaction (and thus the entanglement) is very tiny, so it may not be accessible in simple tabletop experiments. In this direction, for future, one may think about more complicated experimental setups, e.g., including a huge number of particles to enhance the studied effects. This unique feature might be more accessible in curved spacetime backgrounds, such as near black holes or in the cosmological scales. Both of these proposals require more theoretical effort, which could be a way to continue this idea. On the theoretical side, one can relax some assumptions, e.g., by considering test particles with arbitrary spins (like spin-1 particles or quantum rotors \cite{biswas2026gravitonmediatedentanglementlightbending}) or relativistic particles.

	\bibliographystyle{unsrt} 
	\bibliography{ref-GR-spin-spin-SV}

\begin{thebibliography}{10}

\bibitem{PhysRevD.16.953}
K.~S. Stelle.
\newblock Renormalization of higher-derivative quantum gravity.
\newblock {\em Phys. Rev. D}, 16:953--969, Aug 1977.

\bibitem{Polchinski:1998rq}
J.~Polchinski.
\newblock {\em {String theory. Vol. 1: An introduction to the bosonic string}}.
\newblock Cambridge Monographs on Mathematical Physics. Cambridge University
  Press, 12 2007.

\bibitem{Ashtekar:2004eh}
Abhay Ashtekar and Jerzy Lewandowski.
\newblock {Background independent quantum gravity: A Status report}.
\newblock {\em Class. Quant. Grav.}, 21:R53, 2004.

\bibitem{PhysRevLett.119.240402}
C.~Marletto and V.~Vedral.
\newblock Gravitationally induced entanglement between two massive particles is
  sufficient evidence of quantum effects in gravity.
\newblock {\em Phys. Rev. Lett.}, 119:240402, Dec 2017.

\bibitem{PhysRevLett.119.240401}
Sougato Bose, Anupam Mazumdar, Gavin~W. Morley, Hendrik Ulbricht, Marko
  Toro\ifmmode~\check{s}\else \v{s}\fi{}, Mauro Paternostro, Andrew~A. Geraci,
  Peter~F. Barker, M.~S. Kim, and Gerard Milburn.
\newblock Spin entanglement witness for quantum gravity.
\newblock {\em Phys. Rev. Lett.}, 119:240401, Dec 2017.

\bibitem{PhysRevD.108.L101702}
Eduardo Mart\'{\i}n-Mart\'{\i}nez and T.~Rick Perche.
\newblock What gravity mediated entanglement can really tell us about quantum
  gravity.
\newblock {\em Phys. Rev. D}, 108:L101702, Nov 2023.

\bibitem{marletto2024quantuminformationmethodsquantumgravity}
Chiara Marletto and Vlatko Vedral.
\newblock Quantum-information methods for quantum gravity laboratory-based
  tests, 2024.

\bibitem{PhysRevA.101.052110}
Ryan~J. Marshman, Anupam Mazumdar, and Sougato Bose.
\newblock Locality and entanglement in table-top testing of the quantum nature
  of linearized gravity.
\newblock {\em Phys. Rev. A}, 101:052110, May 2020.

\bibitem{Krisnanda_2020}
Tanjung Krisnanda, Guo~Yao Tham, Mauro Paternostro, and Tomasz Paterek.
\newblock Observable quantum entanglement due to gravity.
\newblock {\em npj Quantum Information}, 6(1), January 2020.

\bibitem{Marletto_2018}
Chiara Marletto and Vlatko Vedral.
\newblock When can gravity path-entangle two spatially superposed masses?
\newblock {\em Physical Review D}, 98(4), August 2018.

\bibitem{Bose_2022}
Sougato Bose, Anupam Mazumdar, Martine Schut, and Marko Toroš.
\newblock Mechanism for the quantum natured gravitons to entangle masses.
\newblock {\em Physical Review D}, 105(10), May 2022.

\bibitem{Christodoulou_2023}
Marios Christodoulou, Andrea Di~Biagio, Markus Aspelmeyer, Časlav Brukner,
  Carlo Rovelli, and Richard Howl.
\newblock Locally mediated entanglement in linearized quantum gravity.
\newblock {\em Physical Review Letters}, 130(10), March 2023.

\bibitem{Chakraborty_2023}
Sumanta Chakraborty, Anupam Mazumdar, and Ritapriya Pradhan.
\newblock Distinguishing jordan and einstein frames in gravity through
  entanglement.
\newblock {\em Physical Review D}, 108(12), December 2023.

\bibitem{Beckering_Vinckers_2023}
Ulrich~K. Beckering~Vinckers, Álvaro de~la Cruz-Dombriz, and Anupam Mazumdar.
\newblock Quantum entanglement of masses with nonlocal gravitational
  interaction.
\newblock {\em Physical Review D}, 107(12), 2023.

\bibitem{vedral2022realproblemsquantumgravity}
Vlatko Vedral.
\newblock Are there any real problems with quantum gravity?, 2022.

\bibitem{PhysRevD.102.084015}
G.~P. de~Brito, M.~G. Campos, L.~P.~R. Ospedal, and K.~P.~B. Veiga.
\newblock Quantum corrected gravitational potential beyond monopole-monopole
  interactions.
\newblock {\em Phys. Rev. D}, 102:084015, Oct 2020.

\bibitem{campos2022investigationseffectiveelectromagneticgravitational}
M.~G. Campos, L.~P.~R. Ospedal, and J.~A. Helayël-Neto.
\newblock Investigations on effective electromagnetic and gravitational
  scenarios, 2022.

\bibitem{majumder2026xirphi2nonminimalcoupling}
Avijit~Sen Majumder, Ayan~Kumar Naskar, and Sourav Bhattacharya.
\newblock $\xi r\phi^2$ non-minimal coupling, and the long range gravitational
  potential for different spin fields from 2-2 scattering amplitudes, 2026.

\bibitem{holstein2008spineffectslongrange}
Barry~R. Holstein and Andreas Ross.
\newblock Spin effects in long range gravitational scattering, 2008.

\bibitem{Obukhov_2002}
Yuri~N. Obukhov.
\newblock On gravitational interaction of fermions.
\newblock {\em Fortschritte der Physik}, 50(5-7):711–716, May 2002.

\bibitem{PhysRevD.67.084033}
N.~E.~J. Bjerrum-Bohr, John~F. Donoghue, and Barry~R. Holstein.
\newblock Quantum gravitational corrections to the nonrelativistic scattering
  potential of two masses.
\newblock {\em Phys. Rev. D}, 67:084033, Apr 2003.

\bibitem{PhysRevD.2.1428}
B.~M. Barker and R.~F. O'Connell.
\newblock Derivation of the equations of motion of a gyroscope from the quantum
  theory of gravitation.
\newblock {\em Phys. Rev. D}, 2:1428--1435, Oct 1970.

\bibitem{Barker:1966zz}
Bruce~M. Barker, Suraj~N. Gupta, and Richard~D. Haracz.
\newblock {One-Graviton Exchange Interaction of Elementary Particles}.
\newblock {\em Phys. Rev.}, 149:1027--1032, 1966.

\bibitem{PhysRevD.12.329}
B.~M. Barker and R.~F. O'Connell.
\newblock Gravitational two-body problem with arbitrary masses, spins, and
  quadrupole moments.
\newblock {\em Phys. Rev. D}, 12:329--335, Jul 1975.

\bibitem{Beckering_Vinckers_2025}
Ulrich~K Beckering~Vinckers, Álvaro de~la Cruz-Dombriz, and Anupam Mazumdar.
\newblock Smearing out contact terms in ghost-free infinite derivative quantum
  gravity.
\newblock {\em Classical and Quantum Gravity}, 42(6):065001, February 2025.

\bibitem{Parker:2009uva}
Leonard~E. Parker and D.~Toms.
\newblock {\em {Quantum Field Theory in Curved Spacetime}: {Quantized Field and
  Gravity}}.
\newblock Cambridge Monographs on Mathematical Physics. Cambridge University
  Press, 8 2009.

\bibitem{Birrell:1982ix}
N.~D. Birrell and P.~C.~W. Davies.
\newblock {\em {Quantum Fields in Curved Space}}.
\newblock Cambridge Monographs on Mathematical Physics. Cambridge University
  Press, Cambridge, UK, 1982.

\bibitem{PhysRev.78.29}
Leslie~L. Foldy and Siegfried~A. Wouthuysen.
\newblock On the dirac theory of spin 1/2 particles and its non-relativistic
  limit.
\newblock {\em Phys. Rev.}, 78:29--36, Apr 1950.

\bibitem{PhysRev.111.1011}
Erik Eriksen.
\newblock Foldy-wouthuysen transformation. exact solution with generalization
  to the two-particle problem.
\newblock {\em Phys. Rev.}, 111:1011--1016, Aug 1958.

\bibitem{Silenko_2003}
Alexander~J. Silenko.
\newblock Foldy–wouthuysen transformation for relativistic particles in
  external fields.
\newblock {\em Journal of Mathematical Physics}, 44(7):2952–2966, July 2003.

\bibitem{Murgu_a_2010}
Gabriela Murguía and Alfredo Raya.
\newblock Free form of the foldy–wouthuysen transformation in external
  electromagnetic fields.
\newblock {\em Journal of Physics A: Mathematical and Theoretical},
  43(40):402005, 2010.

\bibitem{Wienczek_2022}
Albert Wienczek, Christopher Moore, and Ulrich~D. Jentschura.
\newblock Foldy-wouthuysen transformation in strong magnetic fields and
  relativistic corrections for quantum cyclotron energy levels.
\newblock {\em Physical Review A}, 106(1), 2022.

\bibitem{PhysRevD.88.084014}
Yuri~N. Obukhov, Alexander~J. Silenko, and Oleg~V. Teryaev.
\newblock Spin in an arbitrary gravitational field.
\newblock {\em Phys. Rev. D}, 88:084014, Oct 2013.

\bibitem{PhysRevD.84.024025}
Yuri~N. Obukhov, Alexander~J. Silenko, and Oleg~V. Teryaev.
\newblock Dirac fermions in strong gravitational fields.
\newblock {\em Phys. Rev. D}, 84:024025, Jul 2011.

\bibitem{PhysRevD.80.064044}
Yuri~N. Obukhov, Alexander~J. Silenko, and Oleg~V. Teryaev.
\newblock Spin dynamics in gravitational fields of rotating bodies and the
  equivalence principle.
\newblock {\em Phys. Rev. D}, 80:064044, Sep 2009.

\bibitem{PhysRevD.76.061101}
Alexander~J. Silenko and Oleg~V. Teryaev.
\newblock Equivalence principle and experimental tests of gravitational spin
  effects.
\newblock {\em Phys. Rev. D}, 76:061101(R), Sep 2007.

\bibitem{PhysRevD.82.104056}
M.~V. Gorbatenko and V.~P. Neznamov.
\newblock Solution of the problem of uniqueness and hermiticity of hamiltonians
  for dirac particles in gravitational fields.
\newblock {\em Phys. Rev. D}, 82:104056, Nov 2010.

\bibitem{alcubierre2025diracequationgeneralrelativity}
Miguel Alcubierre.
\newblock The dirac equation in general relativity and the 3+1 formalism, 2025.

\bibitem{Paper-QED-Spin-Spin}
Radmehr Fathi and Nima Khosravi.
\newblock Spin-spin effects from non-relativistic limit of dirac-pauli-maxwell
  lagrangian (in preparation).

\bibitem{PhysRev.34.553}
G.~Breit.
\newblock The effect of retardation on the interaction of two electrons.
\newblock {\em Phys. Rev.}, 34:553--573, Aug 1929.

\bibitem{PhysRev.39.616}
G.~Breit.
\newblock Dirac's equation and the spin-spin interactions of two electrons.
\newblock {\em Phys. Rev.}, 39:616--624, Feb 1932.

\bibitem{mashhoon2008gravitoelectromagnetismbriefreview}
Bahram Mashhoon.
\newblock Gravitoelectromagnetism: A brief review, 2008.

\bibitem{Meyer_2002}
David~A. Meyer and Nolan~R. Wallach.
\newblock Global entanglement in multiparticle systems.
\newblock {\em Journal of Mathematical Physics}, 43(9):4273–4278, 2002.

\bibitem{biswas2026gravitonmediatedentanglementlightbending}
Dripto Biswas, Sougato Bose, Anupam Mazumdar, and Marko Toroš.
\newblock Graviton-mediated entanglement due to light bending from a quantum
  rotor, 2026.

\end{thebibliography}

	\newpage
	\section*{Appendix: Mathematical Details} \label{app}
	In this appendix, we give the details of the calculations.
	\subsection{The Dirac equation and The Foldy-Wouthuysen  transformation}\label{app-FW}
	We start with the Dirac equation in curved spacetime in the linearized regime, which has the form \cite{PhysRevD.88.084014,PhysRevD.84.024025,PhysRevD.80.064044}:
	\begin{eqnarray}
		g^{00} E \psi = \left[\Phi_g + e^0_{\,\ 0} \gamma^0 m
		+ \frac{1}{2}(\vec{\Gamma} \cdot \vec{\sigma}) + (\vec{\Omega} \cdot \vec{\alpha}) \right] \psi,
	\end{eqnarray}
	where $e^0_{\,\ 0}=1-\frac{h_{00}}{2}$ and we have defined:
	\begin{eqnarray}
		\Gamma^l &=&\frac{1}{2} (\partial_k h_{j0}) \epsilon^{kjl} \\
		\Phi_g &=&  h^i_{\ \ 0} \pi_i - \frac{i}{2} ( \partial_j h^{0j} - \partial_0 h^j_{\ \ j}) \\
		\Omega_l&=&\left( 1 - \frac{h_{00}}{2} \right) \pi_l+ \frac{1}{2} h^i_{\ \ l} \pi_i \nonumber \\
		&-& \frac{i}{2} \partial_l (h_{00}) -\frac{i}{2} (\partial^i h_{il} - \partial_l h^i_{i}).
	\end{eqnarray} 
	Now we will take the non-relativistic limit of the Dirac equation in the linearized regime. The standard procedure for this would be to use the Foldy-Wouthuysen (FW) transformation:
	\begin{eqnarray}
		\mathcal{E} &=& -\frac{h_{0 0}}{2 g^{0 0}}\beta m+\frac{\Phi_g}{g^{00}} + \frac{1}{g^{00}} \vec{\Gamma} \cdot \vec{\sigma} \nonumber \\
		&\simeq&  -\frac{h_{0 0}}{2}\beta m+\Phi_g + \vec{\Gamma} \cdot \vec{\sigma} \\
		\mathcal{G} &=& \vec{\Omega} \cdot \vec{\alpha}.
	\end{eqnarray}
	To proceed in the FW procedure, the matrices should satisfy the following (anti-)commutation relations:
	\begin{eqnarray}
		\mathcal{E} \gamma^0 = \gamma^0 \mathcal{E}, \quad \gamma^0 \mathcal{G} = -\mathcal{G} \gamma^0
	\end{eqnarray}
	which are satisfied in our case. There exists a matrix $S$ which transforms the Hamiltonian to a block-diagonal matrix.
	\begin{eqnarray}
		S = - \frac{i}{2m} \beta \mathcal{G}=-\frac{i}{2 m}(\beta \vec{\Omega} \cdot \vec{\alpha}).
	\end{eqnarray}
	The block-diagonal Hamiltonian, $H'$, takes the following form \cite{PhysRev.111.1011}:
	\begin{eqnarray}\nonumber
		H' &=& H + i \, [S, H] 
		- \frac{1}{2} \, [S, [S, H]] 
		- \frac{i}{6} \, [S, [S, [S, H]]] \nonumber \\
		&+& \frac{1}{24} \, [S, [S, [S, [S, H]]]] + \ldots \nonumber \\
		&-& \dot{S} 
		- \frac{i}{2} \, [S, \dot{S}] 
		+ \frac{1}{6} \, [S, [S, \dot{S}]] + \ldots .
	\end{eqnarray}
	For our Hamiltonian, the block-diagonal Hamiltonian will be:
	\begin{eqnarray}\label{eq:hprime}
		H^{\prime}= \beta \left( m + \frac{1}{2 m} \, \mathcal{G}^2 \right)
		+ \mathcal{E} +\frac{1}{2 m} [\beta \vec{\pi} \cdot \vec{\alpha}, \mathcal{E}].
	\end{eqnarray}
	In this equation, we are keeping things to order one of $h_{\mu \nu}$ and omitting $\mathcal{O}(h^2)$ terms. Since there is no $h_{i j}^{T T}$ in $\mathcal{E}$, we would only deal with the $\mathcal{G}^2$ term in the first-order terms. This leads to the non-relativistic limit of the Dirac equation in the linearized regime for the matter part of the wave function.
	\begin{eqnarray}\label{eq:FW-curved}
		\Rightarrow E \chi &=& \bigg( m (1-\frac{h_{00}}{2}) + \Phi_g + \frac{1}{2}(\vec{\Gamma} \cdot \vec{\sigma}) \nonumber \\
		&+&\frac{(\vec{\Omega} \cdot \vec{\sigma})^2}{2m}+\frac{1}{2 m} [\beta \vec{\pi} \cdot \vec{\alpha}, \mathcal{E}]_{\text{matter}}\bigg) \chi.
	\end{eqnarray}
	We note that $\vec{\Gamma}$ resembles the magnetic field exactly.
	\subsection{Solving the Einstein field equations in the non-relativistic limit}\label{app-fieldeq}
	Now we will calculate $\bar{h}_{i j}$.
	\begin{eqnarray}
		\bar{h}_{ji} &=& -\kappa \int \frac{d^3p}{(2\pi)^3} \frac{1}{k^2} [(2 p_j-k_j) \bar{u}(p-k) \gamma_i u(p) \nonumber \\ 
		&+& (2 p_i-k_i) \bar{u}(p-k) \gamma_j u(p)]
	\end{eqnarray}
	Now, if we take the non-relativistic limit and keep only our desired part (that is, the part that contains both the spin and momentum with different free indices), we would have:
	\begin{eqnarray}
		\tilde{\bar{h}}_{ij} &=& \frac{i \kappa}{2m} \int \frac{d^3p}{(2\pi)^3} \frac{1}{k^2} \chi'^\dagger [ (2 p_j-k_j) (\vec{\sigma} \times \vec{k})_i \nonumber \\
		&+& (2 p_i-k_i) (\vec{\sigma} \times \vec{k})_j ] \chi.
	\end{eqnarray}
	If we go back to the coordinate space, we have
	\begin{eqnarray}
		\bar{h}_{ij}&=&\frac{i \kappa}{2m} \int \frac{d^3k}{(2\pi)^3} \frac{e^{-i\vec{k}\cdot\vec{r}}}{k^2} \biggl[ (2p_j - k_j) \chi'^{\dagger} \sigma^m k^n \epsilon_{mni} \chi  \nonumber\\
		&+& (2p_i - k_i) \chi'^{\dagger} \sigma^m k^n \epsilon_{mnj} \chi \biggr]\nonumber\\
		&=& \frac{-\kappa}{2m} \left(\epsilon_{mni}\partial^n\right) \int \frac{d^3k}{(2\pi)^3} \frac{e^{-i\vec{k}\cdot\vec{r}}}{k^2}  \chi^{\dagger} \sigma^m \pi_j \chi \nonumber\\
		&-& \frac{\kappa}{2m} \left(\epsilon_{mnj}\partial^n \right) \int \frac{d^3k}{(2\pi)^3} \frac{e^{-i\vec{k}\cdot\vec{r}}}{k^2}  \chi^{\dagger} \sigma^m \pi_i \chi \nonumber\\
		&=& \frac{-\kappa}{2m} \epsilon_{mni} \left[ 
		\partial^n \int \frac{\chi^{\dagger} \sigma^m \pi_j \chi}{4\pi r} d^3x \right] \nonumber \\
		&-&\frac{ \kappa}{2m} \epsilon_{mnj} \left[ 
		\partial^n \int \frac{\chi^{\dagger} \sigma^m \pi_i \chi}{4\pi r} d^3x \right].
	\end{eqnarray}
	 Now, if we simplify this expression, we have 
	\begin{eqnarray}
		\overline{h}_{ij} &=& \frac{\kappa}{8\pi m} 
		\int \left( \frac{r^n}{r^3} \right) \biggl[\chi^{\dagger} \sigma^m \pi_j \epsilon_{m n i} \chi \nonumber \\
		&+& \chi^{\dagger} \sigma^m \pi_i \epsilon_{m n j} \chi
		\Biggr] d^3x .
	\end{eqnarray}
	We note that the integration is over $\vec{x}$ which is the location of our source particle and $\vec{r}$ is the difference vector of the location of the particles. What we need to calculate for the FW transformation is:
	\begin{eqnarray}
		\frac{i}{4m} \epsilon_{ijk} \pi^i (h^{lj}) (\sigma_1^k) (\pi_{1l} \chi_1) &=& \frac{1}{4m} \epsilon^{ljk} \partial_l (h_{ij}) (\sigma_1^k) (\pi_1^i \chi_1). \nonumber
	\end{eqnarray}
	So we have
	\begin{eqnarray}
		&& \partial_l (\overline{h}_{ij}) = \frac{\kappa}{8 \pi m}  \Bigg[ \int \chi^{\dagger} \sigma^m \pi_j \epsilon_{mni} \chi \bigg( \frac{r^2 \delta^n_l - 3 r_l r^n}{r^5} \nonumber \\
		&+&\frac{4 \pi}{3} \delta^n_l \delta^3 (r) \bigg) d^3x+ \int \chi^{\dagger} \sigma^m \pi_i \epsilon_{mnj} \chi \bigg( \frac{r^2 \delta^n_l - 3 r_l r^n}{r^5} \nonumber \\
		&+&\frac{4 \pi}{3} \delta^n_l \delta^3 (r) \bigg) d^3x \bigg].
	\end{eqnarray}
	The effect of the second term in the Hamiltonian would be
	\begin{eqnarray}
		&& \frac{-\kappa}{32\pi m^2} \Big( \delta^k_n \delta^l_m - \delta^k_m \delta^l_n \Big) \int \chi_2^{\dagger} \sigma_2^m \pi_{2i} \chi_2 \bigg( \frac{r^2 \delta^n_l - 3 r_l r^n}{r^5} \nonumber \\
		&+&\frac{4 \pi}{3} \delta^n_l \delta^3 (r) \bigg) (\sigma_{1k}) (\pi_1^i \chi_1)d^3x \nonumber\\
		&=& \frac{-\kappa}{32\pi m^2} \bigg[ \int d^3x \ \Big( \chi_2^{\dagger} \sigma_2^l \pi_{2i} \chi_2 \Big) \bigg( \frac{r^2 \delta^n_l - 3 r_l r^n}{r^5} \nonumber \\
		&-&\frac{8 \pi}{3} \delta^n_l \delta^3 (r) \bigg) (\sigma_{1n}) (\pi_1^i \chi_1) \bigg].
	\end{eqnarray}
	Now, if we restore the quantum degrees of freedom of the second particle, we would get:
	\begin{eqnarray}
		&& H_2^{TT} \chi_1 \chi_2=\frac{-\kappa}{32\pi m^2} \biggl[ \frac{1}{r^3} (\vec{\sigma}_1 \cdot \vec{\sigma}_2) (\vec{\pi}_1 \cdot \vec{\pi}_2) \nonumber \\
		&-& \frac{3}{r^5} (\vec{r} \cdot \vec{\sigma}_2) (\vec{r} \cdot \vec{\sigma}_1) (\vec{\pi}_1 \cdot \vec{\pi}_2) \nonumber\\
		&-&\frac{8 \pi}{3} (\vec{\sigma}_1 \cdot \vec{\sigma}_2) (\vec{\pi}_1 \cdot \vec{\pi}_2) \delta^3 (r) \biggr] \chi_1 \chi_2
	\end{eqnarray}
	Now, we apply the same procedure to the first term in the Hamiltonian. We have:
	\begin{eqnarray}
		&& \frac{\kappa}{32\pi m^2} \epsilon_{mni} \epsilon^{ljk} \int \chi_2^{\dagger} \sigma_2^m \pi_{2j} \chi_2 \bigg( \frac{r^2 \delta^n_l - 3 r_l r^n}{r^5} \nonumber \\
		&+&\frac{4 \pi}{3} \delta^n_l \delta^3 (r) \bigg) (\sigma_{1k}) (\pi_1^i \chi_1) d^3x
	\end{eqnarray}
	and then
	\begin{eqnarray}
		&& H_1^{TT} \chi_1 \chi_2= \frac{\kappa}{32 \pi m^2} \biggl[ -\frac{3}{r^5} \Big( \vec{r} \cdot (\vec{\pi}_1 \times \vec{\sigma}_2) \Big) \Big( \vec{r} \cdot (\vec{\pi}_2 \times \vec{\sigma}_1) \Big) \nonumber \\
		&+& \frac{1}{r^3} (\vec{\pi}_1 \times \vec{\sigma}_2) \cdot (\vec{\pi}_2 \times \vec{\sigma}_1) \nonumber \\
		&+&\frac{4 \pi}{3} (\vec{\pi}_1 \times \vec{\sigma}_2) \cdot (\vec{\pi}_2 \times \vec{\sigma}_1) \delta^3 (r) \biggr] \chi_1 \chi_2.
	\end{eqnarray}
	\\
	Both of these terms together give:
	\begin{eqnarray}\label{eq:HTT3}
		&&H^{\text{spin}-2}=\frac{\kappa}{32 \pi m^2}\times\\\nonumber
		&& \left[\frac{-1}{r^3} \Big( 3(\hat{r} \cdot (\vec{\pi}_1 \times \vec{\sigma}_2)) (\hat{r} \cdot (\vec{\pi}_2 \times \vec{\sigma}_1)) - (\vec{\pi}_1 \times \vec{\sigma}_2) \cdot (\vec{\pi}_2 \times \vec{\sigma}_1)\Big) \right. \nonumber\\
		&& \left. +\frac{1}{r^3} \Big(3(\hat{r} \cdot \vec{\sigma}_1) (\hat{r} \cdot \vec{\sigma}_2) (\vec{\pi}_1 \cdot \vec{\pi}_2)-(\vec{\sigma}_1 \cdot \vec{\sigma}_2) (\vec{\pi}_1 \cdot \vec{\pi}_2) \Big) \right. \nonumber \\\nonumber
		&& \left. +\frac{4 \pi}{3} (\vec{\pi}_1 \times \vec{\sigma}_2) \cdot (\vec{\pi}_2 \times \vec{\sigma}_1) \delta^3 (r)+\frac{8 \pi}{3} (\vec{\sigma}_1 \cdot \vec{\sigma}_2) (\vec{\pi}_1 \cdot \vec{\pi}_2) \delta^3 (r)
		\right].
	\end{eqnarray}
	Note that this equation is the same as (\ref{eq:HTT}). This shows that both the scattering amplitude approach and the FW transformation approach yield the same result.
	
\end{document}